\documentclass[traditabstract]{aa} 
\usepackage{color}

\usepackage{natbib}
\usepackage{graphicx}
\usepackage{epsfig}
\usepackage{graphics}
\usepackage[normalem]{ulem}

\newcommand{\nh}{$N_{\rm H}$}

\def\beq{\begin{equation}}
\def\enq{\end{equation}}
\def\begar{\begin{eqnarray}}
\def\endar{\end{eqnarray}}
\newcommand{\Msun}{\mbox{$M_{\odot}\;$}}
\def\lsim{\;\raise0.3ex\hbox{$<$\kern-0.75em\raise-1.1ex\hbox{$\sim$}}\;}
\def\gsim{\;\raise0.3ex\hbox{$>$\kern-0.75em\raise-1.1ex\hbox{$\sim$}}\;}

\def\beq{\begin{equation}}
\def\enq{\end{equation}}
\def\begar{\begin{eqnarray}}
\def\endar{\end{eqnarray}}
\def\mathnew{\mathsurround=0pt}
\def\simov#1#2{\lower .5pt\vbox{\baselineskip0pt \lineskip-.5pt
        \ialign{$\mathnew#1\hfil##\hfil$\crcr#2\crcr\sim\crcr}}}

\def\cms{\rm ~cm^{-2}}
\def\kms{\rm ~km~s^{-1}}
\def\ergs{\rm ~erg~s^{-1}}

\def\cms{\rm ~cm^{-2}}

\def\kms{\rm ~km~s^{-1}}
\def\mum{\rm ~\mu m}

\def\ergs{\rm ~erg~s^{-1}}

\def\enf{\rm ~erg~cm$^{-2}$~s$^{-1}$}

\def \chan {{\it Chandra}}
\def \xmm {{\it XMM-Newton}}

\def\src{Kes~69~}

\def \hcm {\hbox {\ifmmode $ atom cm$^{-2}\else atom cm$^{-2}$\fi}}
\def \arcmin {\hbox{$^\prime$} }
\def \arcsec {\hbox{$^{\prime\prime}$} }

\def\approxgt{\mathrel{\hbox{\rlap{\lower.55ex \hbox {$\sim$}}
        \kern-.3em \raise.4ex \hbox{$>$}}}}
\def\approxlt{\mathrel{\hbox{\rlap{\lower.55ex \hbox {$\sim$}}
        \kern-.3em \raise.4ex \hbox{$<$}}}}
\def\mathnew{\mathsurround=0pt}

\titlerunning{Hard X-ray sources in Kes~69}
\authorrunning{F.~Bocchino et al.}

\begin{document}

\title{
A population of isolated hard X-ray sources near the supernova remnant Kes 69}

\author{F.~Bocchino\inst{1}, A.M.~Bykov\inst{2},
Y.~Chen\inst{3,4},
A.M.~Krassilchtchikov\inst{2,5}, K.P.~Levenfish\inst{2},
M.~Miceli\inst{6,1}, G.G.~Pavlov\inst{7,5},
Yu.A.~Uvarov\inst{2,5},
X.~Zhou\inst{8}
}
\institute{INAF-Osservatorio Astronomico di Palermo, Piazza
del Parlamento 1, 90134 Palermo, Italy; bocchino@astropa.unipa.it
\and
A.F. Ioffe Institute for Physics and Technology,
St.\ Petersburg, Russia, 194021; byk@astro.ioffe.ru
\and
Department of Astronomy, Nanjing University, Nanjing, 210093, China
\and
Key Laboratory of Modern Astronomy and Astrophysics (Nanjing University), Ministry of Education, China
\and
St.\ Petersburg State Polytechnical University, St.\ Petersburg,
Russia, 195251
\and
Dipartimento di Fisica, Universit\`a di Palermo, Piazza del Parlamento 1, 90134 Palermo, Italy
\and
525 Davey Laboratory, Pennsylvania State
University, University Park, PA 16802; pavlov@astro.psu.edu
\and
Purple Mountain Observatory, CAS, 2 West Beijing Road, Nanjing 210008, China
}

\abstract{
 {Recent X-ray observations of the supernova remnant IC443 interacting with molecular clouds have shown the presence of a new population of hard X-ray sources related to the remnant itself, which has been interpreted in terms of fast ejecta fragment propagating inside the dense environment. Prompted by these studies, we have obtained a deep {\sl XMM-Newton} observation of the supernova remnant (SNR) Kes 69, which also shows signs of shock-cloud interaction.
We report on the detection of 18 hard X-ray sources in the field of
Kes~69, a significant excess of the expected galactic source population in the field,
spatially correlated with CO emission from the cloud in the remnant environment.
The spectra of 3 of the 18 sources can be described as hard power laws with photon
index $<$ 2 plus line emission associated to K-shell transitions.
We discuss the two most promising scenarios for the interpretation of the sources, namely fast ejecta fragments (as in IC443) and cataclysmic variables. While most of the observational evidences are consistent with the former interpretation, we cannot rule out the latter.}
}


\maketitle

\section{Introduction}

 {
The high-resolution X-ray imaging of supernova remnants (SNRs) interacting with dense ambient matter has shown that a population of previously unknown compact sources are sometimes present in the field of view, visible at high X-ray energies (above $\sim 2$ keV). These are the X-ray
shrapnels -- the ballistically moving isolated fragments of
ejected material radiating both thermal and non-thermal X-rays
\citep{bbp05,bku08}. 
}

Ejecta fragments have been long known (\citealt{aet95,mta01,lh03}), but when they reach a dense cloud, they may become bright and compact sources in
IR (\citealt{bku08}) and X-ray, with luminosity $L_X >$ 10$^{31} \ergs$ \citep[][]{b02,b03}.
Since the fragment lifetime in the cloud is about a few hundred years,
a few of such isolated sources are expected to be visible in the field of
an SNR entering a molecular cloud.
The shrapnel-type sources produce multicomponent X-ray spectra consisting of:
(i) a relatively faint thermal continuum with $kT\la 1$ keV,
(ii) a power-law nonthermal continuum with photon index $\sim1.5$, and
(iii) non-thermal line emission due to K-shell ionization produced by the
intense flux of electrons (accelerated at the bow-shock) that collide
with a metal-rich ejecta fragment.

 {The ballistically moving clumps are potentially an important source of information about the explosive supernova (SN) event. In fact, they presumably correspond to initial ejecta inhomogeneities, located in the faster outer layers, and their chemical composition may be indicative of the nucleosynthesis processes which occurred at early stages of SN evolution. Therefore, they provide complementary information with respect to the rest of X-ray emitting ejecta.
It should be noted, however, that a firm detection of an ejecta fragment
is a challenging task, because fragment's lifetime is only a few hundred
years, and the brighter is the fragment, the shorter is the lifetime.
}

SNRs interacting with molecular clouds are the natural sites to
study the physics of fast fragments. One of the best cases is
IC~443, which emits bright molecular lines of OH, CO, and H$_2$,
excited by a shock. IC~443 is an extended object (about 45$\arcmin$
in size at a distance of about 1.5 kpc) with the brightest GeV
gamma-ray emitting SNR shell \citep{aaa10b}. It is also a source
of TeV gamma-rays \citep{aaa07}. Hard X-ray sources in IC~443
related to shocked clumps have been detected by \citet{bb00} and
further studied by \cite{bb03}, \citet{bbp05}, \citet{bkk08}
and \citet{bku08}.

 {
Motivated by the results obtained in the SNR IC~443, we have started an observational campaign aimed at detection of fast fragments in SNRs interacting with molecular clouds. The SNR G21.8--0.6 (Kes~69) is a good candidate where such sources may be seen. }
It is an extended incomplete radio shell of about
20$\arcmin$ in size \citep[e.g.,][]{sg70} at an estimated distance of
about 5.2 kpc \citep[e.g.,][]{tl08,zcs09}. The Imaging
Proportional Counter aboard the {\sl Einstein Observatory} was used
by \citet{s90} to obtain a noisy X-ray image of Kes~69. Using
{\sl ROSAT} PSPC observations, \citet[][]{ywr03} estimated an absorbing
column of \nh = 2.4$\times$10$^{22} \cms$.

{\sl Spitzer} observations of \src by \citet{rrt06} and
\citet{hra09} have
revealed bright molecular emission lines of OH, CO, and H$_2$,
excited by a shock. In particular, the SNR contains a filament
along the bright southern radio shell of $\sim$ 15$^\prime$ extension
seen with {\sl Spitzer} IRAC in the 4.5$\mum$ band and attributed to
the molecular hydrogen emission line, as well as an extended
OH (1720 MHz) emission region that indicates the presence of a
molecular shock \citep[see][ and references therein]{hra09}.
Millimeter band observations of CO and HCO+ lines towards \src
performed by \citet[][]{zcs09} provided strong evidence for the
association between \src\ and the $\sim$~+85~$\kms$ component of
molecular gas. 


  {The established interaction of \src with a molecular cloud
makes this SNR a promising target for a search of shrapnel X-ray
sources. Since the size of the remnant (about 20$\arcmin$) is
comparable with the field of view of \xmm\ EPIC cameras, we
conducted \xmm\ observations of the field to study the faint population of hard compact X-ray sources and look for the signature of fast ejecta fragments. Here we report the detection of 18 sources in the field of \src, a significant excess with respect to the expected Galactic field sources of the same flux.
We have found some of these sources to emit hard power law spectra with possible 
K$_{\alpha}$ X-ray lines of Si, Fe, Ca, and Ti. Such spectra support
the suggested shrapnel nature of the sources.}

\section{{\sl XMM-Newton} observations and data analysis}

\label{xmm}

The field of \src was observed with the \xmm\  observatory (\citealt{jla01}) on 2009 October 8
(ObsID~0605480101, PI:~M.~Miceli)
for 60 ksec with the EPIC-MOS camera (\citealt{taa01}) in the Full Frame Mode and with the EPIC-PN
camera (\citealt{sdb01}) in the Extended Full Frame Mode
(the medium filter was used for all the cameras).
In the course of \xmm\ data reduction, we selected EPIC events with FLAG=0,
patterns 0--4 for EPIC-PN camera, and patterns 0--12 for EPIC-MOS cameras.

To filter out periods of enhanced particle background,
the original event lists were screened by using the sigma-clipping algorithm of the ESAS software\footnote{http://xmm.esac.esa.int/pub/xmm-esas/xmm-esas.pdf}.
After the filtering, the net good time exposure of the field was reduced to about 54 ks.
SAS v9.0.0 was used for EPIC data processing, and the HEASOFT~6.3 suite,
including XSPEC v.12.3.1, was used for spectral fitting.

\section{Results}

\subsection{Diffuse emission}

Hard (3--10 keV) and soft (1--3 keV) X-ray maps of \src are shown
in Figure~\ref{xmmimage}. Diffuse X-ray emission is clearly seen on
the latter map. The diffuse emission region is bordered by the radio
emission contours obtained from the 1.4~GHz NVSS survey \citep{ccg98}.
{No significant diffuse emission is found on the hard X-ray map, while several point-like sources are present. We verified that the brightest fluctuation of the residual diffuse emission in the hard X-ray map is at the level of only $3\sigma$ above the background. The faint fluctuations generally correlates with the bright soft X-ray emission, so we argue that the very faint hard diffuse emission is the tail of the thermal emission visible in the soft X-ray map, and we do not consider it any further.}

 {
Spectral analysis of the diffuse extended emission of 
a mixed morphology SNR requires a thorough modeling of multiple
components, both thermal (with possibly overionized plasma) and  
nonthermal. Such a detailed analysis of the diffuse soft 
X-ray emission of \src is beyond the scope of this work.
However, to estimate the absorption column toward this remnant and to compare it to the absorption of the compact sources, a
}
spectrum of the extended diffuse emission region in the southern shell
(shown as black rectangle in the middle panel of Figure~\ref{xmmimage})
has been extracted and fitted with the absorbed thermal plasma {\it mekal} model.
For the spectral fitting, a channel binning with at least 15 counts per bin was used. 
The fitting yielded $N_{\rm H}$ = (2.8$\pm$0.4)$\times$10$^{22}$~cm$^{-2}$,
$T$~=~(0.8$\pm$0.2)~keV (errors are at the 99\% confidence level).
The absorption column depth is consistent with that derived
previously from the {\sl Einstein} and {\sl ROSAT} observations. 
The higher temperature in the 
{\sl ROSAT} fit derived by \citet{ywr03} is most likely due to the contribution of the point sources 
unresolved by {\sl ROSAT} in the wide region analyzed.
The postshock temperature $T$~$\approx$~0.8~keV corresponds to the estimated age 
of about 10,000 years, assuming the Sedov stage solution. 
The forward shock velocity can be estimated as $\sim$ 1,000 $\kms$. 
The velocities of the isolated ejecta fragments in the vicinity of the shell 
should be comparable to the shock velocity.

\subsection{Hard X-ray sources detection and spectroscopy}

\begin{figure}

     \centerline{\includegraphics[width=7.0cm]{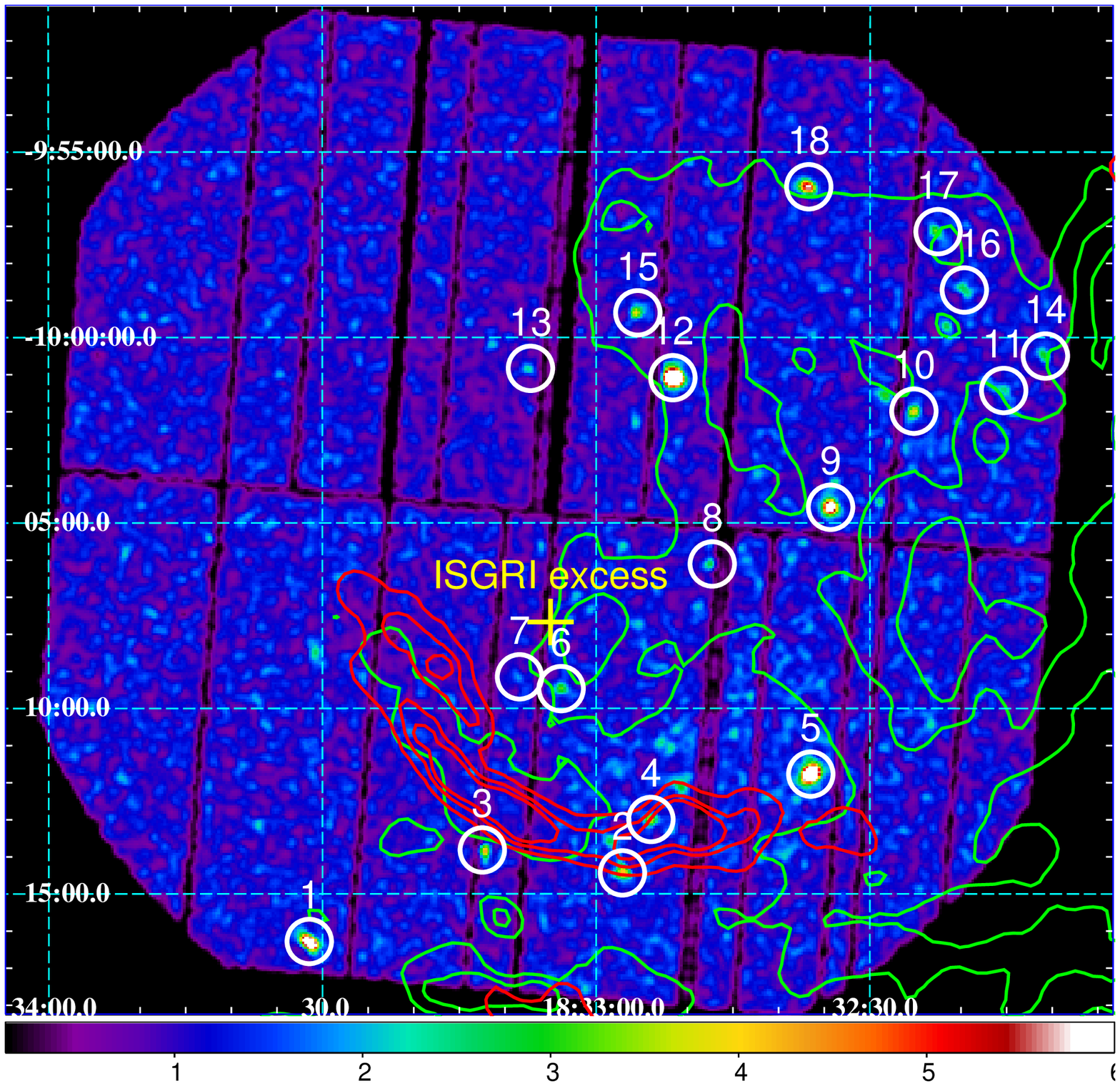}}
     
     \centerline{\includegraphics[width=7.0cm]{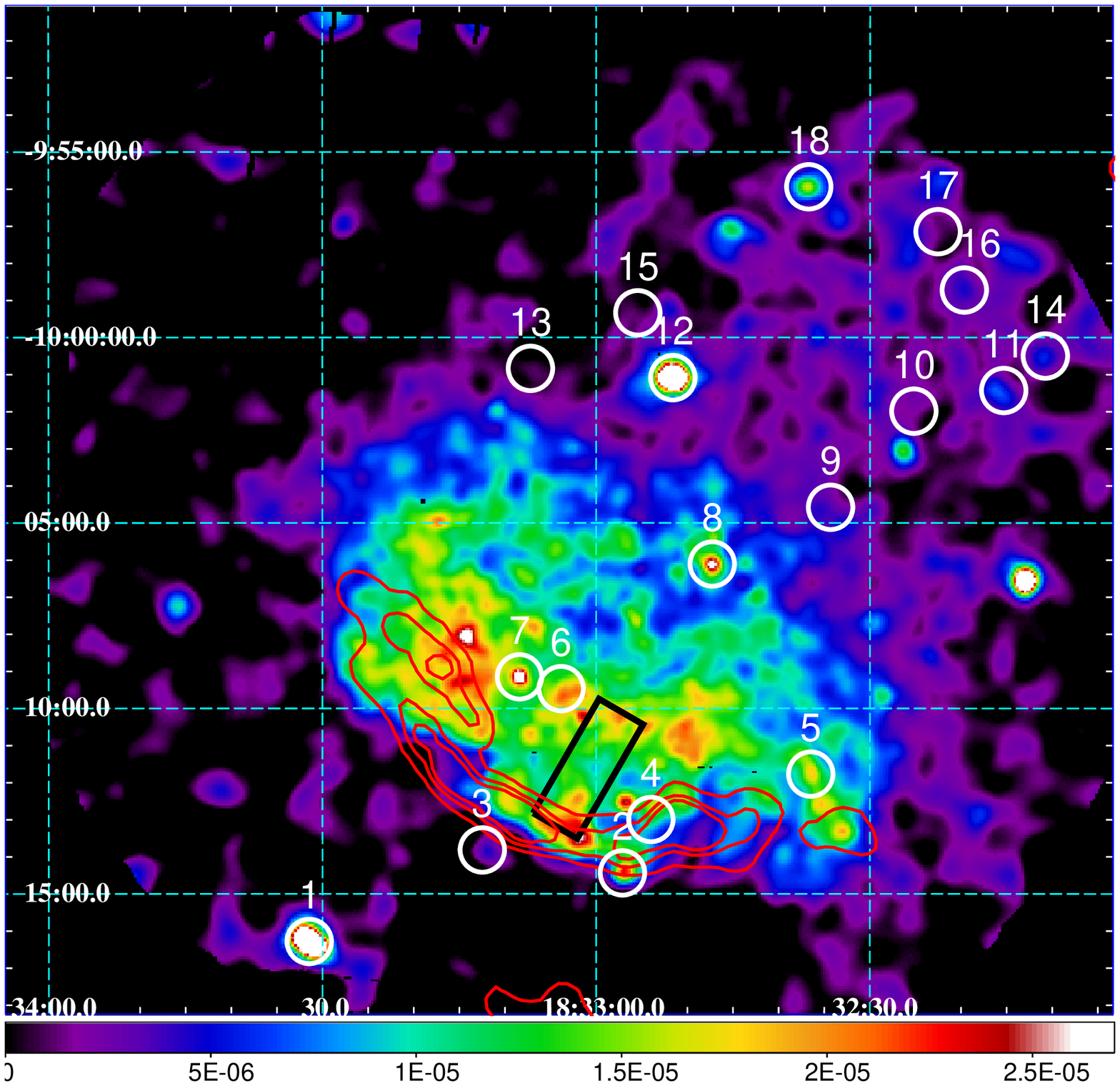}}

     \centerline{\includegraphics[width=7.0cm]{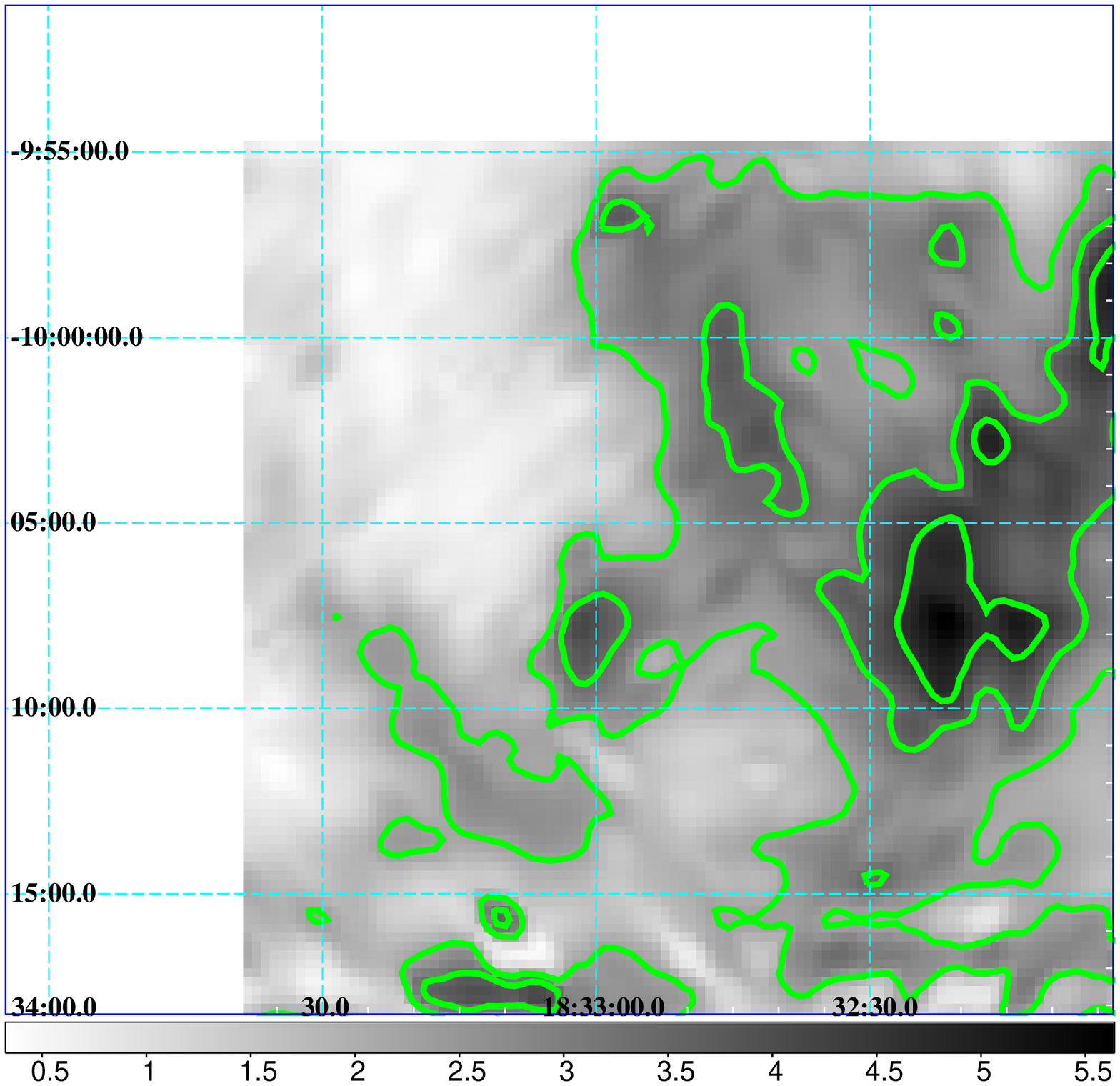}}
\caption{\emph{Upper panel}: EPIC PN count map of \src obtained
in the 3--10~keV band. The bin size is 4\hbox{$^{\prime\prime}$}.
The white circles indicate point-like sources detected in this image;
the red contours denote intensity of the 1.4~GHz emission from the NVSS survey \citep{ccg98}; the green contours denote the
$^{12}$CO (J = 1 -- 0) intensity in the 80-81 km s${^-1}$ velocity range (see lower panel). The yellow cross denotes the position
of the hard X-ray excess seen with {\sl INTEGRAL} ISGRI. 
\emph{Middle panel}: Adaptively smoothed (to a signal-to-noise ratio of 16)
and vignetting-corrected EPIC count-rate image (MOS-equivalent counts
per second per bin) of \src in the 1--3~keV band. The black rectangle is the region used for the spectrum of the diffuse emission. The superimposed hard
X-ray sources and the radio contours are the same as in the upper panel.
\emph{Lower panel:} $^{12}$CO (J = 1 -- 0) intensity map in the velocity
interval 80--81 km/s (linearly interpolated to a resolution
of 0\farcm24 \citep[see][]{zcs09}. The contour levels are at
40\%, 60\%, and 80\% of the maximum.
}
\label{xmmimage}
\end{figure}

\begin{figure}

\centerline{\includegraphics[width=8.8cm]{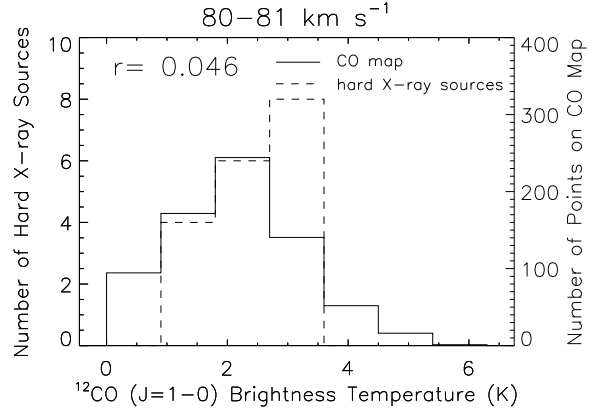}}
\caption{Histogram of brightness temperature values in the $^12$CO maps in the 80-81 km s$^{-1}$ velocity range (Zhou et al. 2009, also visible in the lower panel of Fig. \protect\ref{xmmimage}) which shows the molecular cloud interacting with the remnant (solid line). Overimposed, the histogram of the same temperatures at the location of the 18 hard X-ray sources detected in the XMM-Newton observation (dashed line). 
The $r$ value is the Pearson correlation coefficient, indicating a positive correlation (see appendix A for details). 
The source peak is offset toward higher value with respect to the cloud mean value, indicating that X-ray sources cluster at brighter-than-average CO map points.}

\label{histo80}
\end{figure}

The SAS task edetect\_chain was applied to the 3--10 keV image to
detect and study the population of point-like sources 
in the field of \src and to identify the sources likely associated with
this remnant.
The thermal emission from the SNR makes almost no contribution
in the energy band chosen.
Eighteen sources, whose properties are
listed in Table~\ref{listsrc}, have been detected with the
likelihood above 30 $\sigma$ (marked in Figure~\ref{xmmimage}). The
 {
positions of the hard X-ray sources are remarkably correlated with the $^12$CO
emission at the velocity associated with the SNR ($80-81$ km s$^{-1}$,
\citealt{zcs09}), whose image and contours are shown in
Figure~\ref{xmmimage}. The correlation is better seen in Figure \ref{histo80}, 
which shows that the average brightness CO temperature at the location of the 18 sources 
is higher than the average cloud brightness temperature, which means that the hard X-ray 
sources tend to be preferentially found in dense regions indicated by bright CO emission (see Appendix A for details about this method).
The general correlation between the source position and the
molecular cloud that is
interacting with the remnant makes the sources good
candidates for X-ray emitting ejecta
fragments. 
}

 \begin{table*}
 \footnotesize
 \center
 \caption{List of sources detected in the 3--10 keV band in
 the field of view of the XMM-Newton observation of the \src SNR
 \label{listsrc}}
 \medskip
 \setlength{\tabcolsep}{1mm}
  {
 \centering\begin{minipage}{18.0cm}
 \begin{tabular}{cccccccc} \hline \hline
 Src & Name & Count-rate & $N_H$ & $\Gamma$ & $\chi^2/dof$ & $10^{-14} F_{2-10\rm{\ keV}}$ & Possible association \\ 
 & & PN cnt ks$^{-1}$ & $10^{22}$ cm$^{-2}$ & & & erg cm$^{-2}$ s$^{-1}$ & in SIMBAD within $15^{\prime\prime}$ \\ \hline
 1 &  XMMU J183331.4-101616 & $9.8\pm 2.0$ & 0.7$_{-0.2}^{+0.3}$ & 1.2$_{-0.2}^{+0.2}$ & 1.1  & 23$_{-12}^{+4}$ &               \\
 2 &  XMMU J183257.1-101426 & $8.1\pm 0.8$ & 2.6$_{-1.0}^{+1.6}$ & 2.3$_{-0.5}^{+0.9}$ & 0.82 & 4.4$_{-3.3}^{+1.4}$ & IRAS 18302-1016 ($13\arcsec$) \\
 3 &  XMMU J183312.4-101349 & $10.6\pm 0.9$ & 3.6$_{-1.1}^{+2.1}$ & 0.6$_{-2.8}^{+2.8}$ & 0.47 & 2.5$_{-1.3}^{+1.3}$ & 1XMM J183312.5-101351 (1\arcsec)  \\
 4 &  XMMU J183254.0-101259 & $4.2\pm 0.6$ & 1.7$_{-0.9}^{+1.9}$ & 1.7$_{-0.8}^{+1.0}$ & 0.82 & 2.9$_{-2.2}^{+1.1}$ &               \\
 5 &  XMMU J183236.4-101146 & $35.3\pm 2.3$ & 3.0$_{-1.0}^{+1.0}$ & 0.6$_{-0.2}^{+0.3}$ & 1.1  & 16$_{-15}^{+4}$ &     XGPS-I J183236-101144 ($3\arcsec$) \\
 6 &  XMMU J183303.8-100928 & $4.0\pm 0.5$ & 1.2$^{+1.9}_{-1.0}$ & 1.0$_{-0.9}^{+1.2}$ & 1.2  & 2.2$_{-1.5}^{+0.9}$ &               \\
 7 &  XMMU J183308.4-100909 & $3.2\pm 0.5$ & 0.5$_{-0.3}^{+0.5}$ & 1.7$_{-0.7}^{+0.8}$ & 0.84 & 1.1$_{-0.8}^{+0.6}$ &               \\
 8 &  XMMU J183247.2-100606 & $4.3\pm 0.6$ & 0.6$_{-0.2}^{+0.6}$ & 1.6$_{-0.7}^{+0.8}$ & 1.2  & 1.8$_{-1.3}^{+0.6}$ &               \\
 9 &  XMMU J183234.3-100434 & $10.9\pm 1.1$ & 45$_{-25}^{+33}$ & 4.0$_{-3.0}^{+3.0}$ & 1.1  & 5.7$_{-5.7}^{+1.3}$ & 2MASS J18323431-1004360 ($1\arcsec$)
               \footnote{$m_J$ = 16.1.}  \\
 10 &  XMMU J183225.2-100159 & $6.2\pm 0.7$ & 21$_{-18}^{+42}$ & 2.8$_{-3.1}^{+5.9}$ & 0.86 & 3.4$_{-3.39}^{+0.9}$ &     XGPS-I J183225-100158 ($0.5\arcsec$) \\
 11 &  XMMU J183215.3-100125 & $5.7\pm 0.8$ & $<3.5$ & 1.1$_{-1.5}^{+2.6}$ & 0.57  & 2.6$_{-2.5}^{+1.9}$ &               \\
 12 &  XMMU J183251.6-100105 & $5.8\pm 1.4$ & 0.7$_{-0.1}^{+0.1}$ & 1.25$_{-0.1}^{+0.1}$& 0.97 & 26$_{-5}^{+3}$ & XGPS-I J183251-100106 ($1\arcsec$)
               \footnote{An $m_R$=21.6 star, according to \citet{mwc10}.}  \\
 13 &  XMMU J183307.2-100050 & $3.0\pm 0.5$ & 0.8$_{-0.5}^{+1.7}$ & 1.1$_{-0.9}^{+4.8}$ & 0.42  & 1.4$_{-0.8}^{+1.0}$ &               \\
 14 &  XMMU J183210.8-100029 & $9.2\pm 1.1$ & $<9.0$ & 0.2$_{-0.8}^{+3.1}$\footnote{1 $\sigma$ errors.} & 1.35  & 3.3$_{-2.7}^{+1.7}$ & XGPS-I J183210-100031 ($1\arcsec$) \\
 15 &  XMMU J183255.4-095920 & $6.3\pm 0.7$ & 11$_{-9}^{+21}$ & 1.2$_{-1.9}^{+2.8}$ & 1.4  & 4.2$_{-4.1}^{+1.4}$ &               \\
 16 &  XMMU J183219.6-095843 & $8.0\pm 1.0$ & 8.4$_{-6.2}^{+14.6}$ & 2.5$_{-2.1}^{+3.1}$ & 0.79  & 3.2$_{-3.0}^{+1.1}$ &               \\
 17 &  XMMU J183222.5-095708 & $9.0\pm 1.0$ & 15$_{-11}^{+37}$ & 1.6$_{-1.8}^{+4.9}$ & 1.1  & 4.7$_{-4.7}^{+1.4}$ &               \\
 18 &  XMMU J183236.7-095556 & $20.0\pm 1.4$ & $<0.7$ & 0.2$_{-0.5}^{+0.4}$ & 1.0  & 11$_{-6.5}^{+3.5}$ &               \\ \hline
\end{tabular} \\
\end{minipage}
}
\end{table*}

 {
We have extracted the EPIC spectra for all the sources
rebinned to achieve a signal-to-noise
ratio $>$ 4 per bin, and fitted simultaneously
both MOS and PN spectra. The ancillary response files were
produced with the SAS ARFGEN task, and the EVIGWEIGHT task
(\citealt{ana01}) was used to correct for vignetting effects. An
empty area in the northeastern part of the image was used as a
background region.
Parameters of power law fits to the spectra 
of the point-like hard sources are listed in Table~\ref{listsrc}.
}

Src~3, 5 and 18 were selected for a detailed study
based on the possible presence of X-ray line emission lines in the spectrum similar to those expected in fast moving ejecta knot according to \citet{b02}.
The lines in Src 3 and 5 can be
interpreted as Fe line complex and as traces of the 1.8~keV Si line.
The observed line features at 3.5~keV and 4.7~keV in Src~13 could 
originate from $^{40}$Ca and $^{48}$Ti, 
respectively.

\begin{figure}
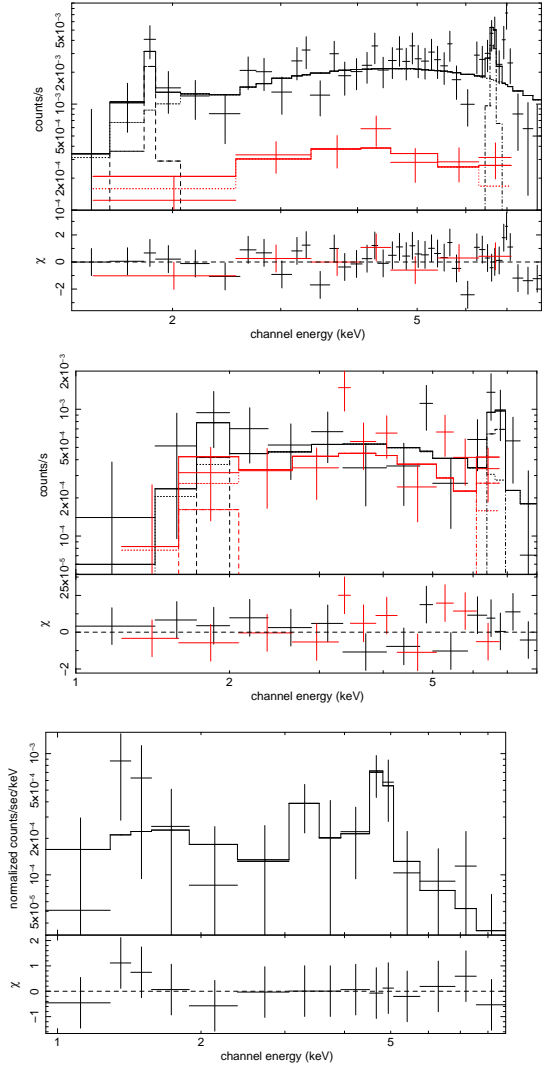


\centerline{\includegraphics[width=0.5\columnwidth, angle=270]{f2a.eps}}

\centerline{\includegraphics[width=0.53\columnwidth, angle=270]{f2b.eps}}

\centerline{\includegraphics[width=0.56\columnwidth, angle=270]{f2c.eps}}

\caption{\xmm\ EPIC-PN (black) and MOS2 (red) spectra of Src~5 ({\it
upper panel}), Src~3 ({\it middle panel}), and Src~13 ({\it lower panel}) 
with best-fit models (power law plus two narrow gaussians) and residuals. 
}
\label{xmmspectra}
\end{figure}

%
The spatial count distributions of Src~3, 5, and 13 are consistent
with that of a point source. It should be noted here that
with the few arcsec resolution of \xmm\ the upper limit
of a point-like source extension at the estimated distance
of 5.2 kpc is about 10$^{17}$ cm.

Two circular regions of 16\arcsec radius centered at Src~3 
and Src~5 were chosen for spectral analysis. Unfortunately, both sources
were projected onto the damaged MOS1 CCD6 chip; therefore
only PN and MOS2 events were taken into account.
For spectral analysis of Src~13 a circular region 
of 20\arcsec radius containing 160 PN counts was chosen.
A 10 counts per bin PN spectrum was used to locate the positions
of line features in the spectrum.

\subsubsection{Src 5}

Src~5 was first detected by \citet{hww04} in the \emph{XMM-Newton}
Galactic Plane Survey (XGPS). This source is brighter than Src~3 and Src~13, and its spectrum
is shown in the upper panel of Figure~\ref{xmmspectra}. A simple absorbed power-law model
provides $\chi^2=63.9$ (at 46 dof), but it does not fit the bright line-like features
visible at about 1.8~keV and 6.5~keV. The quality of the fit improves significantly by adding
two narrow gaussians to describe the lines\footnote{Since the absorbing column
\nh=3.8$^{+1.2}_{-1.5}\times 10^{22}$ cm$^{-2}$ is consistent with that found for
the whole remnant, we fixed it here to $4\times10^{22}$ cm$^{-2}$ to obtain more restricted
estimations of the other model parameters.}.
We found that the line energies are $1.84^{+0.06}_{-0.06}$ keV and $6.64^{+0.06}_{-0.08}$ keV, respectively,
the photon index is $\Gamma=0.6^{+0.3}_{-0.2}$, and the improved 
$\chi^2$ is 47.8 (at 43 dof).
Confidence contours for the line normalization coefficients and line energies are
shown in the upper panels of Figure~\ref{confid}.
The line features can be associated with K-shell transitions of Si (at $1.85$ keV)
and Fe (at $6.64$ keV) originating from a metal-rich ejecta knot.
Assuming that the source is at the same distance of 5.2 kpc as Kes~69, the line
luminosities are $\sim$~6$\times$10$^{39}$ photons~s$^{-1}$ for the Si line,
and $\sim$ 3$\times$10$^{39}$ photons~s$^{-1}$ for the Fe line (with relatively
large uncertainties, as shown in Figure~\ref{confid}, upper panels).
 {Both these values
and the ratio of the Si and Fe line luminosities are in
agreement with those predicted for fast ($\sim 2700$ km s$^{-1}$) and small (size of $3\times 10^{16}$ cm) ejecta knots having a mass of $10^{-3}$ $M_\odot$ moving inside a dense ($\sim 10^3$ cm$^{-3}$) molecular cloud 
(see Table~1 in \citealt{b02}, where luminosities for a fragment are {shown}).}

We investigate a possible
thermal origin of Src~5 by fitting its spectrum with an absorbed
thermal model. Although the probable Si line is not well described by this model,
the quality of the fit is still acceptable ($\chi^2=52.3$ at 46 dof).
The best-fit parameter values for the 
thermal fit are $kT$=~8.8$^{+3.7}_{-2.1}$~keV and
\nh=(7.8$^{+2.4}_{-1.6}$)$\times$10$^{22}$~cm$^{-2}$. Since in the
thermal scenario the column density is higher than that found for
Kes~69, a lower limit for the distance to Src~5 is 5.2 kpc (i.e. the
distance to the remnant), and its X-ray luminosity is $>$10$^{33}$
erg/s. These extreme values can be indicative of a very energetic
coronal flare, as X-ray luminosities up to $\sim$10$^{33}$ erg~s$^{-1}$ and
temperatures up to 10 keV have been observed in the most energetic
flares of active stars (e.g., \citealt{ffr05}). Another possible
source of such a high temperature and luminosity could be a
cataclysmic variable star (CV). Nevertheless, we point out
that by comparing the source count-rate in our observation with that
observed by \citet{hww04} in the XGPS observations (17.1$\pm$0.7 PN
counts/ks and 19.3$\pm$3.8 PN counts/ks, for Src~3 and Src~5,
respectively, in the 0.4--6 keV band), we found that the source
luminosity is consistent with being constant. Also, the hardness
ratio of 0.75 obtained by \citet{hww04} is very similar to
0.70$\pm$0.05 obtained in our new observation. The stationarity of
the X-ray luminosity and hardness ratio both concur in making a coronal 
flaring or a CV origin of Src~5 unlikely, though still not excluded (see also
Section~\ref{IR}).

\subsubsection{Src 3}

The background-subtracted spectrum of Src~3 is shown in
the middle panel of Figure~\ref{xmmspectra}. This spectrum is 
substantially absorbed (model \nh\ values are about 10$^{22}$~cm$^{-2}$ 
-- consistent with that of the soft diffuse emission from the remnant), 
and contains line-like features at about 1.8 keV and 6.7 keV. Like in Src~5, we
fixed the \nh\ to 4$\times$10$^{22}$~cm$^{-2}$, and found that a
power law model with $\Gamma$ = 1.3$^{+0.5}_{-0.7}$ and two narrow 
gaussians\footnote{Because of the
poorer statistics, we fixed the line energies to those found for Src~5.}
describes the spectra significantly better than a simple power law
model ($\chi^2$=20.4 at 21 dof, and $\chi^2$=28.5 at 22
dof, respectively). 
 {
Assuming that Src~3 is at 5.2 kpc, the Fe
line luminosity is (1.6$\pm$0.9)$\times$10$^{39}$~photons~s$^{-1}$ and the
Si line luminosity is $<6\times$10$^{39}$~photons~s$^{-1}$.
The best-fit values are in good agreement with
those predicted by \citet{b02} for a supersonic ejecta fragment propagating in a molecular cloud, the only difference from Src 5 being a smaller best-fit velocity ($\sim 1000$ km s$^{-1}$).
}
We investigated a possible thermal origin of Src~3 
by fitting its spectrum with an absorbed 
thermal model and found that, with such a model, we can only constrain 
the temperature to be above 7 keV.

\subsubsection{Src 13}
\begin{figure}
\centerline{\hbox{
\includegraphics[width=3.5cm, angle=90]{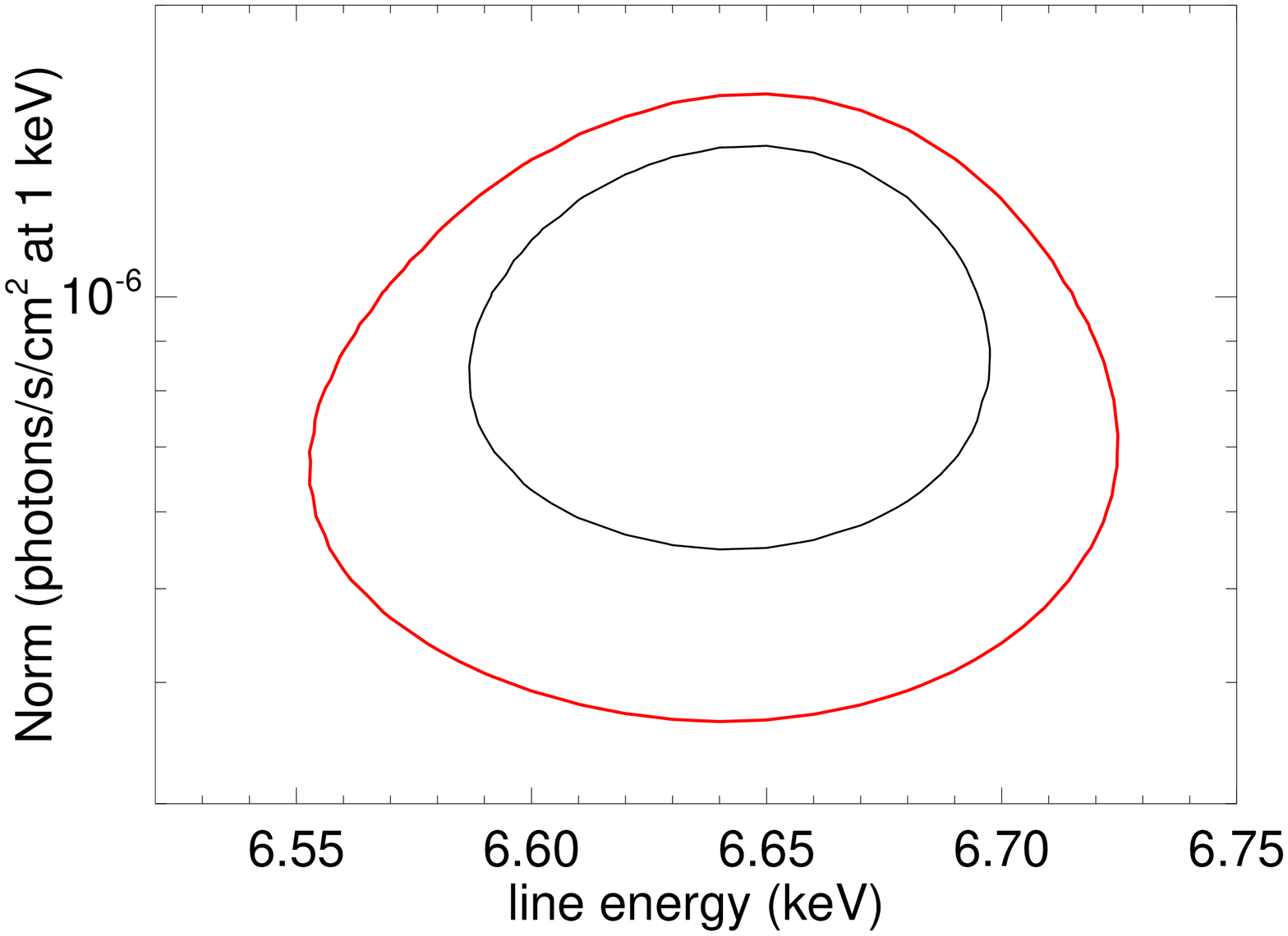}
\includegraphics[width=3.0cm, angle=90]{f3c.eps}
}}
\centerline{\hbox{
  \includegraphics[width=3.5cm, angle=90]{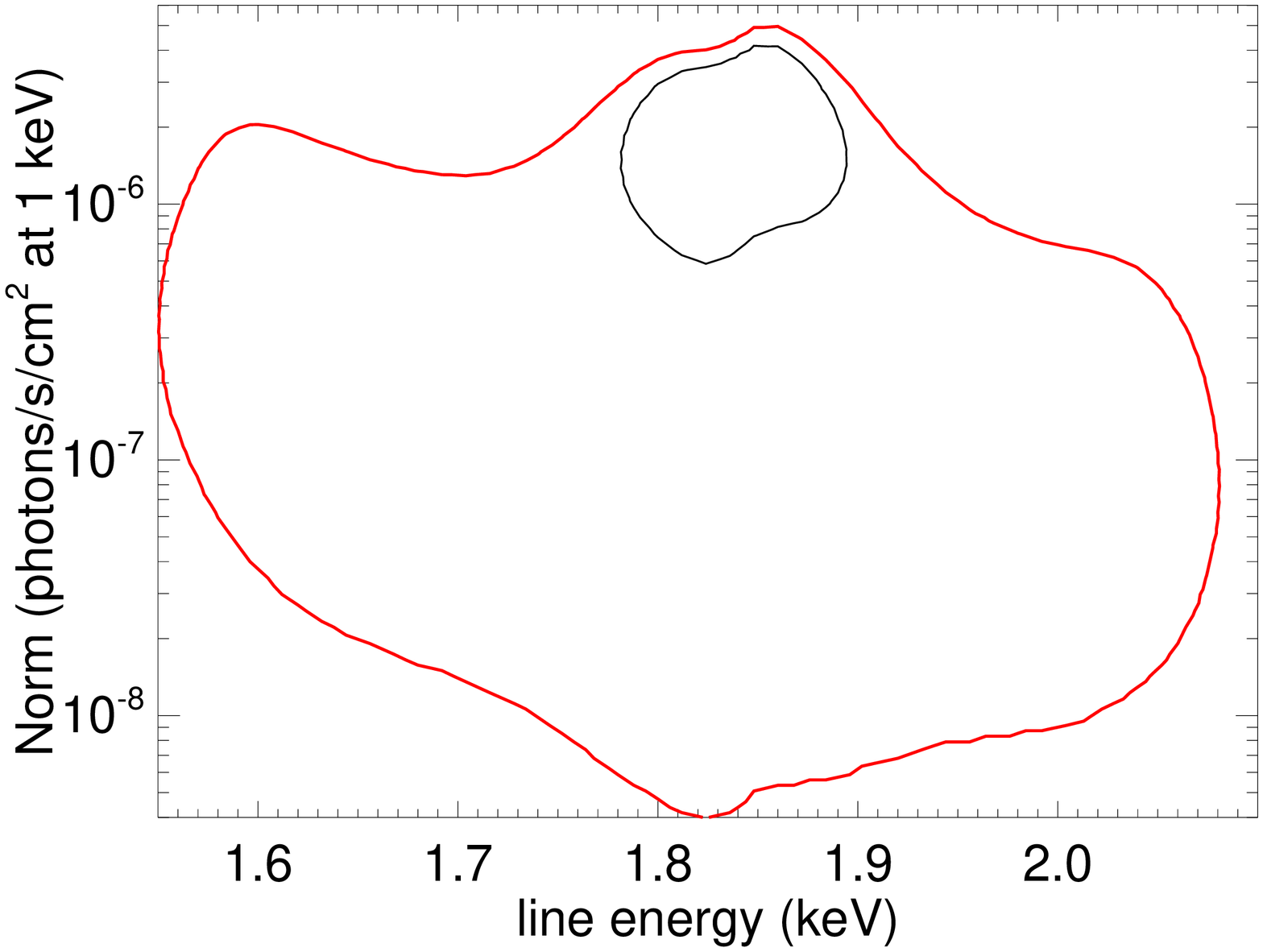}
  \includegraphics[width=3.0cm, angle=90]{f3d.eps}
}}  
  \caption{Confidence contours of the line centroid and normalization
for the Fe (\emph{upper left}) and Si (\emph{lower left}) lines in Src~5, 
and for the $^{48}$Ti (\emph{upper right}) and Ca (\emph{lower right}) lines in Src~13.
The contours correspond to the 68\% and 90\% confidence levels.
}
\label{confid}
\end{figure}

A background-subtracted spectrum of Src~13 is shown in
the lower panel of Figure~\ref{xmmspectra}. The spectrum
contains line-like features at about 3.5~keV and 4.7~keV.
Confidence contours for the line normalization coefficients 
and line energies are shown in the lower panels of Figure~\ref{confid}.
A power law model with $\Gamma$ = 1.1$^{+4.8}_{-0.9}$ and 
two narrow gaussians at fixed energies fits the spectrum ($\chi^2$=11.0 at 31 dof). 
The line luminosities at 5.2 kpc are about 7.3$\times$10$^{38}$ photons~s$^{-1}$
and 1.7$\times$10$^{39}$ photons~s$^{-1}$ for the 3.5~keV and 4.7~keV lines,
respectively, in agreement with the range of values predicted 
for an ejecta fragment in a molecular cloud. Unlike the spectra
of Src 3 and 5, the spectrum of Src 13 does not show Fe complex
line features, a fit 
with a thermal model is much worse than the non-thermal one.

\begin{table*}
\footnotesize
\center
\caption{Fluxes of possible infrared counterparts to Src~3, 5, and 13. Infrared fluxes of Src {\sl o51, o31, o32, o13} are quoted
for {\sl Spitzer} IRAC bands I1, I2, I3, I4, {\sl Spitzer} MIPS
band M1, and 2MASS bands J, H, K$_s$ (see Figure~\ref{ir1}).
The flux units are 10$^{-13}$~erg~cm$^{-2}$~s$^{-1}$.
Errors are at the 68\% level.
Estimates of extinction factors (i.e. the value $10^{A_{band}/2.5}$, $A_V$ is computed from the X-ray absorption) at positions of Src~3 and 5
are listed as {\sl ext5} and {\sl ext3}.}
\label{tab_ir}
\setlength{\tabcolsep}{0.5mm}
\begin{tabular}{ccccccccc} \hline \hline
Band   & I1 & I2 & I3 & I4 & M1 & J & H & K$_s$ \\  
$\lambda, \mu$m	& 3.6 & 4.5 & 5.8 & 8.0 & 24 & 1.2 & 1.7 & 2.2 \\ \hline
Source &   &     &     &     &      &   &    &           \\
o51   & 2.2$\pm$0.1 & 1.5$\pm$0.1 & $<$1.6 & $<$1.6 & $<$0.9 & $\cdots$ & $\cdots$ & $\cdots$ \\
o31   & 4.5$\pm$0.1 & 2.4$\pm$0.1 & $<$1.2 & $<$1.4 & $<$2.7 & $\cdots$ & 3.8$\pm$0.4 & 4.5$\pm$0.3 \\
o32   & 210$\pm$0.9 & 120$\pm$0.4 & 91.0$\pm$0.8 & 54.1$\pm$0.4  & 6.3$\pm$0.1 & 6.3$\pm$0.2 & 75$\pm$1.6 & 140$\pm$2.7 \\ 
o13   & 5.4$\pm$0.4 & 2.9$\pm$0.3 & 2.0$\pm$0.3 & 1.5$\pm$0.3  & $\cdots$ & 2.2$\pm$0.2 & 6.6$\pm$0.2 & $<$5.9  \\ \hline
%
{\sl ext5}  & 4.0 & 3.5 & 3.2 & 2.8 & 1.0 & 575 & 58 & 13 \\
{\sl ext3}  & 2.5 & 2.4 & 2.2 & 2.0 & 1.0 & 76 & 16 & 5.8 \\
\hline \hline
\end{tabular}
\end{table*}

\section{Hard X-ray {\sl INTEGRAL} ISGRI Data of Kes 69}

The field of \src was a target of Galactic plane scans
with the {\sl INTEGRAL} gamma-ray observatory. The archival
data obtained with the ISGRI camera during the scans
comprise about 1700 ks of fully coded field of view observations
in the years 2002  -- 2010.

A weak hard X-ray source is marginally seen in the central part
of the remnant (see the upper panel of Figure~\ref{xmmimage}).
The source position is close to the projected center of 
an imaginary triangle with vertices in Src 3, 5, and 13.
The HEAVENS\footnote{http://www.isdc.unige.ch/heavens/}
survey \citep{wrm10}
provides 3$\sigma$ detections in the 18~--~50~keV and
50~--~150~keV bands (see the lower right panel of Figure~\ref{xmmspectra}).
To estimate the hard X-ray flux of the sources,
intercalibration with a bright nearby SNR G21.5$-$0.9
\citep[e.g.,][]{krl07} was performed
to yield 5.7$\times$10$^{-12}$\enf\ in the 18~--~60~keV band,
and 1.2$\times$10$^{-12}$\enf\ in the 60~--~120~keV band.

The highest plasma
temperatures consistent with the {\sl XMM-Newton} data, discussed in
\S3, are too low to yield ISGRI detection above 60
keV. Thus, the detection of the INTEGRAL source associated with \src is
evidence in favor of a non-thermal origin of the emission from
{(at least some of) the compact sources of this remnant. The power-law models for Src 3, 5, and 13
obtained from the \xmm\ data would yield (1--5)$\times$ 10$^{-12}$ erg
cm$^{-2}$ s$^{-1}$ in the 18--60 keV band, which is compatible with the flux detected by ISGRI in the same band.
The extrapolation of the power-law model of the three sources to the 60--120 keV band would yield a flux higher than the flux observed by ISGRI in that band. Therefore, if the sources were indeed the X-ray counterpart of the ISGRI emission, a spectral break would be required around 60 keV.
}

\section{IR imaging and photometry of \src}
\label{IR}

The field of Kes~69 was the target of {\sl Spitzer} MIPS
observations {r15602176},
{r15584768}, {r15621120},
{r15634432}, and {r15627008},
performed on 2005 Sep 30 (PI: S.\ Carey) and {\sl Spitzer}
IRAC observations {r12103680}, 
{r12107264}, and {r12110592}, performed on 2004
Oct 07 (PI: E.\ Churchwell).

The standard MOPEX~18.4.9 software \citep{mrk06}
was used to construct mosaic images and detect point sources
from the archival BCD-level data pre-processed by the S18.7.0
(IRAC) and S18.12.0 (MIPS) pipelines. The net exposure
of the mosaic maps was equal to 2--5 frames of 1.2~s (IRAC)
and 10--18 frames of 2.62~s (MIPS).

\begin{figure}
\includegraphics[width=0.89\columnwidth]{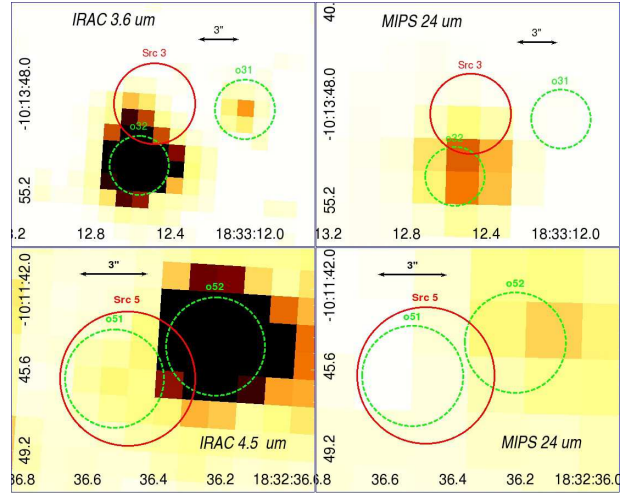}
\caption{Environments of Src~3 ({\em top panels}) and Src~5 ({\em bottom panels}) as seen by {\sl Spitzer} IRAC camera at 3.6 and 4.5 $\mu$m, and MIPS at 24 $\mu$m.
The sources Src~3 and Src~5 detected by {\sl XMM-Newton} are shown as
solid red circles, while the IR sources are shown as
dashed green circles, both with labels with names referred to in the text.}
\label{ir1}
\end{figure}

\begin{figure}
\includegraphics[width=0.98\columnwidth]{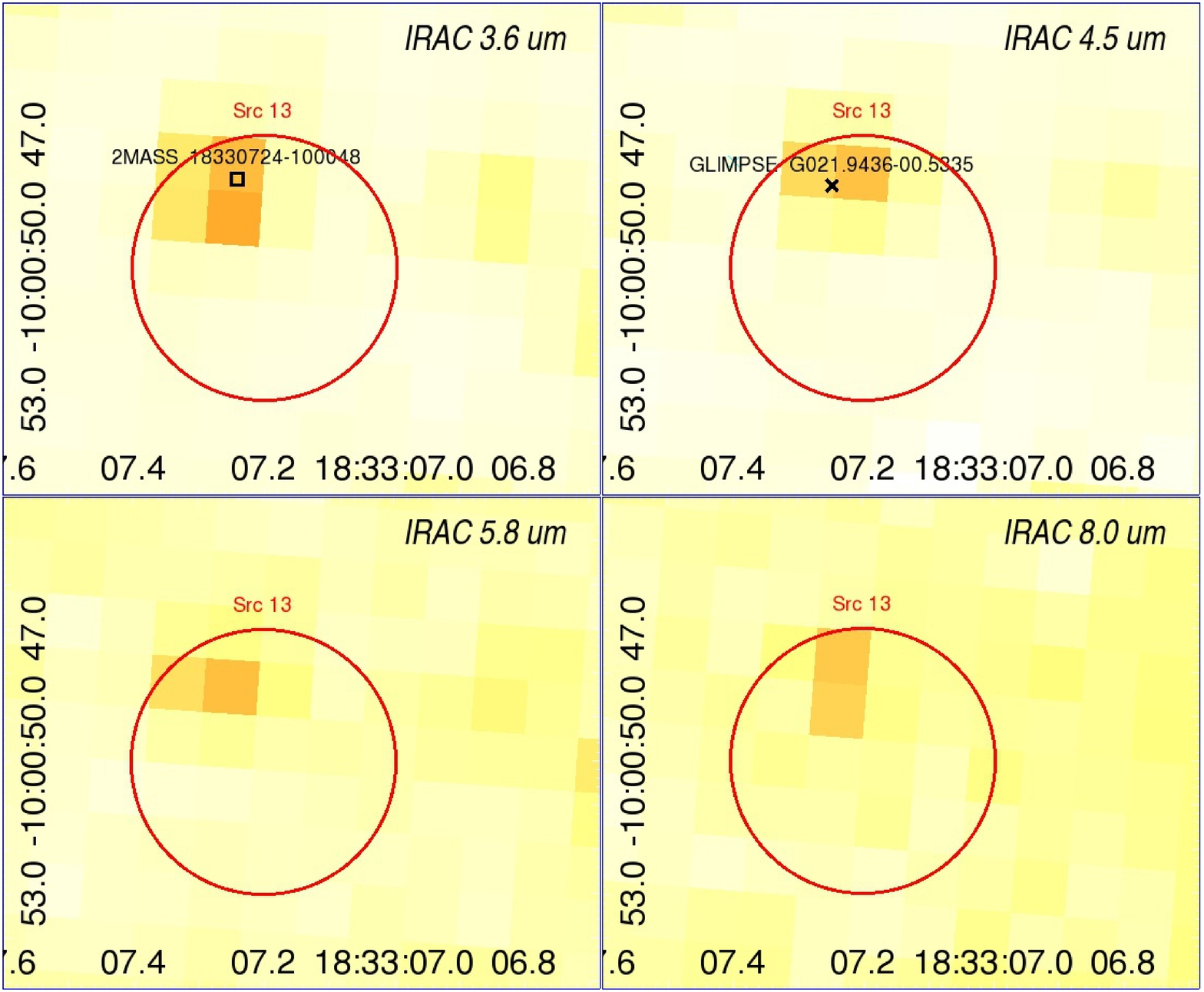}
\caption{Close environments of Src~13 as seen with {\sl Spitzer} IRAC at 3.6, 4.5, 5.8 and 8.0 $\mu$m. The red circle marks the position of the {\sl XMM-Newton} detected source.
}
\label{ir2}
\end{figure}

Two point-like sources were found in the immediate vicinity
of Src~3 (see the upper panels of Figure~\ref{ir1}) at
$\alpha = 18^{\rm h}33^{\rm m}12\fs 0,\ \delta = -10^\circ 13' 49$\arcsec
(source {\sl o31}, $6.2\arcsec$ from the X-ray source) and at
$\alpha = 18^{\rm h}33^{\rm m}12\fs 6,\ \delta = -10^\circ 13' 53$\arcsec
(source {\sl o32}, at $4.7\arcsec$).
Two point-like sources were found in the immediate vicinity
of Src~5 (see the lower panel of Figure~\ref{ir1}) at
$\alpha = 18^{\rm h}32^{\rm m}32\fs 2,\ \delta = -10^\circ 11' 45$\arcsec
(source {\sl o51}, at $<1\arcsec$ from the X-ray source) and at
$\alpha = 18^{\rm h}32^{\rm m}36\fs 5,\ \delta = -10^\circ 11' 46$\arcsec
(source {\sl o52}, at $4.3\arcsec$). All these positions are in the J2000 reference frame.
Another point-like source is clearly seen at about 2\arcsec from Src~13
(see Figure~\ref{ir2}) at
$\alpha = 18^{\rm h}33^{\rm m}07\fs 2,\ \delta = -10^\circ 00' 48$\arcsec
(source {\sl o13}).

The source {\sl o52}, listed in the USNO-B1.0 catalogue \citep{mlc03},
has {\sl B} = 17.7, {\sl R} = 14.1, which means that it is most likely 
a foreground star. The other four
sources can be treated as the possible infrared counterparts of Src~3, 5, and 13.
Medium IR fluxes of these sources measured with MOPEX and near IR fluxes listed
in the 2MASS survey \citep{scs06} are summarized in Table~\ref{tab_ir} along with extinction factors.

 {
As no specific data can be found in the available catalogs to
estimate the distances and velocities or somehow indicate the
nature of the IR sources that are possibly connected
with Src~3, 5, and 13, it is useful to compare the IR colors of
Src {\sl o51, o31, o32,} and {\sl o13 } with those typical for various classes
of objects, paying a particular attention to ejecta fragment candidates
and CVs, which have been pointed out as plausible interpretations
of Src~3, 5 and 13.
}

The F$_{\rm I1}$/F$_{\rm I3}$ and F$_{\rm I2}$/F$_{\rm I4}$ flux ratios of
Src {\sl o51, o31, o32,} and {\sl o13 } are larger than those in Figure~2
of \citet{rrt06}, which contains examples of IRAC flux
ratios of shock-excited zones in SNRs.
The m$_{\rm I1}$-m$_{\rm I2}$ and m$_{\rm I2}$-m$_{\rm I3}$ colors of
Src {\sl o51, o31, o32, o13 } are well within the range of
IRAC colors reported by \citet{sra06} and
\citet{gas06} for faint X-ray sources in the
Galactic Center region, which are usually interpreted as accreting binaries.
The IR spectral energy distribution of a CV source AE Aqr reported by
\citet{dth07} is similar to that of Src~{\sl o32}, but
24$\mu$m flux of Src~{\sl o32} rescaled to the distance of AE Aqr
(102 pc) appears much lower.
Thus, the assumption that the potential IR counterparts
of Src~3 and 5 are CVs seems unlikely, although there is also no
firm evidence for detection of an ejecta fragment in the IR bands.
However, it should be noted that the $\sim$10$^{34}$ erg~s$^{-1}$
IR luminosities of the possible counterparts of Src 3, 5, and 13 
are within the range predicted for IR emission of ejecta fragments
interacting with molecular clouds \citep[see e.g.][]{bku08}. 

\section{Discussion}

\subsection{Interpretation as ejecta fragments}

The apparent correlation of the location of the hard X-ray sources
detected in the field of \src with the CO emission of molecular
material associated with the remnant together with consistent
values of absorbing columns obtained from X-ray spectra of the
sources naturally leads to an SNR-related interpretation. Moreover, there is
an apparent excess of the sources with 2--10 keV fluxes above 
(2--3)$\times$10$^{-14}$ erg cm$^{-2}$ s$^{-1}$ within the region of about
15\hbox{$^\prime$}$\times$20\arcmin size where the SNR is
interacting with the molecular cloud (shown in Figure~\ref{xmmimage}).
The $\log(N)-\log(S)$ statistics of X-ray sources in the \xmm\ Galactic
plane survey by \citet{mwc10} predicts about 40 sources of 2--10
keV flux above 3$\times$10$^{-14}$\enf\ per square degree. Therefore,
only about 3--4 galactic field sources are expected in the region on the
statistical ground. The X-ray luminosities of the sources 
are $L_X \gsim$10$^{31}\ergs$
if they are located at the estimated distance of Kes~69.

 {If the source are related to \src, then isolated ``shrapnels" or ``knots", associated with fast moving ejecta
fragments, form an interesting explanation for a population of hard X-ray sources in the field of SNRs. A massive
individual fragment moving supersonically through a molecular cloud
would have a luminosity $L_X \gsim$ 10$^{31} \ergs$ in the 1--10 keV
band and would be observable with \xmm\ and \chan\ from a few kpc
distance.
Such fragments have indeed been detected in hard X-rays inside the IC~443
SNR (\citealt{bkk08}, \citealt{bku08}, and references therein), and
the X-ray sources in Kes~69, which is also interacting with
molecular material (\citealt{zcs09}), can also be interpreted in
the same way.
}

The model of fast supernova fragments predicts two X-ray emission
components \citep{b03}. The first one is thermal X-ray emission from
a hot shocked ambient gas behind the fragment bow shock, with the
spectrum of an optically thin thermal shocked plasma of an ISM cloud
abundance. The second emission component is nonthermal. Interaction
of the fast electrons (accelerated at the fragment bow-shock) with
the fragment body produces both hard continuum and X-ray and IR line
emission, including the K-shell lines of the Si and Fe group
elements \citep{b02}\footnote{{As for the radio emission, the electrons accelerated at the fragment bow shock in most cases are in the MeV regime, well below GeV. So synchrotron radio emission is not expected to be observable for small fragments. However, an HII region may be created due to gas ionization by a powerful soft X-ray -- far UV emission of the fragment bow shock.
\citet{bku08} made some simple modeling of the radio continuum from such an HII region that is expected to be produced by a fast ejecta fragment moving through dense ambient matter. The scale size of the radio emitting HII region is  below arcmin if the distance to Kes 69 is about 5 kpc and its flux at 1.4 GHz is likely about 30 mJy (scaling from the IC443 case).}}.

As {mentioned} in Sect. 3.2, the upper limit of a point-like source extension 
at the estimated distance of 5.2 kpc is about 10$^{17}$ cm.
Individual compact fragments of scale sizes below $\sim$10$^{17}$~cm 
would be optically thick to the K-shell X-ray line
absorption if the metal content in the source exceeds
about 10$^{-4}$~\Msun.
Indeed, for Li-like to F-like ions the probability of auto-ionization 
by a K$_{\alpha}$ photon (i.e., the resonant Auger destruction) 
is significantly larger than that of photon re-emission 
\citep[see, e.g.,][]{l05}, resulting in the true
line photon absorption. The line optical depth $\tau_X$ with respect to the
resonant absorption of a K$_{\alpha}$ line of energy $E_X$ (in keV) for an
element $X$ (of atomic weight $A$) can be estimated as
\begin{equation} \label{depth}
\tau_X \simeq 3\cdot {f_{lu} \over 0.5} \cdot {\xi_{X^i} \over 0.4} \cdot A^{0.5} \cdot
\mbox{N}_{X,17} \left[ T_6 + A \cdot (2.3\times10^3
\beta_{\sigma_v})^2 \right]^{-0.5} E_X^{-1},
\end{equation}
where $\xi_{X^i}$ is the relative abundance of the ion $i$ of the element $X$,
$\mbox{N}_{X,17}$ is  the column density of the element $X$ in the source along the line of sight
in units of 10$^{17} \cms$, and $f_{lu}$ is the relative
line oscillator strength of the K$_{\alpha}$ line.
The characteristic line width is assumed to be equal to the 
Doppler full width at half maximum, in which both the thermal ion 
velocity for temperature $T$ = 10$^6 \cdot T_6$ K and the micro-turbulent 
velocity ($\beta_{\sigma_v} = \sigma_v / c$)  are accounted for 
\citep[see, e.g.,][]{nfm99}. 

This implies, in particular, that the line fluxes of such 
fragments are expected to be about 10$^{31} \ergs$ for most of the
X-ray lines, and that, for example, $^{40}$Ca and $^{48}$Ti line
fluxes are expected to be comparable even though the $^{40}$Ca
abundance typically exceeds that of $^{48}$Ti in supernova
ejecta (line saturation effect). The maximal model yields of
$^{48}$Ti in core-collapsed SNe range from about 4$\times$10$^{-5}$~\Msun 
\citep{ww95} to 2$\times$10$^{-4}$~\Msun
\citep{tnh96}, and are subject to rather substantial model
uncertainties. Although core-collapsed SNe are naturally associated
with molecular cloud environments {(see e.g. \citealt{jcw10}) and reference therein}, still SN~Ia type origin can not
be excluded for Kes~69, since no compact remnant has been found there
yet. According to the models of SN~Ia nucleosynthesis
developed by \citet{mrp10}, the yield of $^{48}$Ti can exceed
10$^{-3}$ \Msun in SN~Ia ejecta. Therefore, one may conclude that
some ejecta fragments can show line emission of $^{48}$Ti at the
level estimated at 68\% significance in Src~13 without any signs of
$^{44}$Ti, which is expected to decay at the estimated age of Kes~69.
 {
In conclusion, both the location, the X-ray spectra and the infrared counterpart of Src 3, 5, and 13 are consistent
with the expectations for supernova ejecta fragments interacting
with a dense ambient medium.
}

\subsection{Alternative explanations}

 {
The X-ray spectra of Src 3,5 and 13 may also be fitted with a very hot thermal component. We have therefore explored other explanations for the detected sample of hard X-ray sources.
Accreting dwarfs and stars with active coronae can be sources of
hard X-ray emission \citep[e.g.][]{fpr76,cmn81,pr85} and are therefore an interesting possibility.
}
The X-ray luminosity of some CVs can reach up to $L_X \sim
10^{34}~\ergs$ \citep[see e.g.][]{she11}, but the majority
of CVs are much fainter. Integrated emission of CVs,
especially of intermediate polars, is considered to be the
dominating component in both the Galactic Center \citep[see
e.g.][]{mbb09} and Galactic Ridge emission \citep[see
e.g.][]{srg06,rsc09}. Individual spectra of magnetic
CVs, obtained with {\sl ASCA} by \citet{ei99}, show
the Fe $K_{\alpha}$ emission lines, both the fluorescent emission at $\sim$
6.4 keV and from highly ionized Fe ions. The spectra can be attributed to
the hot postshock plasma emission behind the accretion shock
standing above the white dwarf surface. X-ray reflection from the
white dwarf surface can account for the observed  fluorescent Fe
$K_{\alpha}$ line. Hard X-rays in the 17--60 keV band have been detected so far by
{\sl INTEGRAL} IBIS \citep[see][]{lll03,uld03} from 37
CVs \citep[see, e.g.,][]{ktr10}. Thus, CVs could be
alternative candidate sources for Src 3 and 5, but
not for Src~13, due to the putative presence of Ca and $^{48}$Ti lines 
along with the absence of Fe lines
in its spectrum. However, while the ejecta
shrapnel model can account for the emission in the 60~--~120~keV
band with the flux of about 1.2$\times$10$^{-12}$\enf\ detected by
{\sl INTEGRAL} ISGRI from Kes~69 field, the CVs most likely can
not. Since the ejecta shrapnels
are expected to be extended, dedicated \chan\
observations can constrain the spatial extensions of Src~3, 5, and
13 to distinguish between the two interpretations.

\section{Summary and conclusions}

 {
We have performed a study of the supernova remnant \src\  to search for compact hard X-ray sources. Such studies, carried out in remnants which are interacting with molecular clouds, may reveal a population of fast ejecta fragments, which may provide useful hints for the remnant progenitors. We have detected 18 sources in the 3-10 keV band, about 3 times the expected number of galactic sources in this area of the sky. Moreover, the position of the source are not random, but correlate with the CO emission of the cloud interacting with the remnant. We have selected 3 of the sources for further spectral analysis on the basis of the brightness and visual inspection of their spectra (Src 3, 5 and 13 in Table \ref{listsrc}). The spectra can be fitted with a non-thermal model plus a few emission lines corresponding to K-shell transition of Si and Fe. We have also analyzed the Spitzer IRAC and MIPS image of the field, finding possible IR counterparts for Src 3, 5 and 13. The X-ray spectra and IR flux ratio are consistent with a model of a fast ejecta fragment propagating inside the molecular cloud, making \src\  the second SNR for which an evidence exists for this new population of hard X-ray sources.
}

 {
Uncertainties in the X-ray modeling makes the alternative explanation in terms of galactic cataclysmic variables also possible, but the excess with respect of $\log (N) - \log (S)$ counts, the location of the sources, the detection of the remnant by INTEGRAL/IBIS and some of the IR flux ratios make this interpretation less probable then the other. Further data are needed to say the final word about the nature of the sources.
}

\begin{acknowledgements}

This study has been supported by Ministry of Education and Science
of Russian Federation (contract 11.G34.31.0001)
and by the RFBR grants 11-02-12082-ofi-m-2011 and 11-02-00253. 
Support from P-21 and OFN-16 programs of RAS is acknowledged as well.
AMK and YuAU acknowledge support from scientific school
NSh-4035.2012.2.
GGP acknowledges partial support from NASA grant NNX09AC84G.
YC acknowledges support from the NSFC grant 10725312 and the 973
Program grant 2009CB824800.
This work has been partially supported by ASI-INAF 
agreement n. I/009/10/0.

\end{acknowledgements}


\bibliographystyle{aa}
\bibliography{ms10}

\appendix
\section{A simple method to assess the spatial correlation between a set of point-like source and the CO diffuse emission maps}

Fig. \ref{xmmimage}, upper panel, shows the detected hard X-ray sources and the $^{12}$CO contours emission in the 80-81 km s$^{-1}$ range, which is the range of velocity associated with the Kes 69 remnant (\citealt{zcs09}). By eye, the source seem to be correlated with the cloud emission, but one need to assess this correlation in more detail. A simple way to do this is described below.

We have computed the histogram of the 80-81 km s$^{-1}$ $^{12}$CO brightness temperature map shown in Fig. \ref{xmmimage}, lower panel. This histogram is plotted as a solid line in Fig. \ref{histo80} and has a peak around $2-2.5$ K. We have also computed the brightness temperature at the locations of the 18 detected hard X-ray source in the same map, and we also computed the histogram and overplotted it as a dashed line in the Fig. \ref{histo80}. We stress that in case of a uniform distribution of the hard X-ray sources, their histogram should match the cloud histogram. If the sources tend to be located preferentially outside the cloud, the peak of their histogram should lie to the left of the peak of the cloud. Conversely, if the sources tend to be inside the cloud, their peak should fall at a temperature higher than the cloud temperature, which means that the source have a tendency to be located in regions of high brightness temperature. This is exactly what happens in case of the histograms computed using the $80-81$ km s$^{-1}$ $^{12}$CO brightness temperature map (Fig. \ref{histo80}).

\begin{figure}

\centerline{\includegraphics[width=6.75cm]{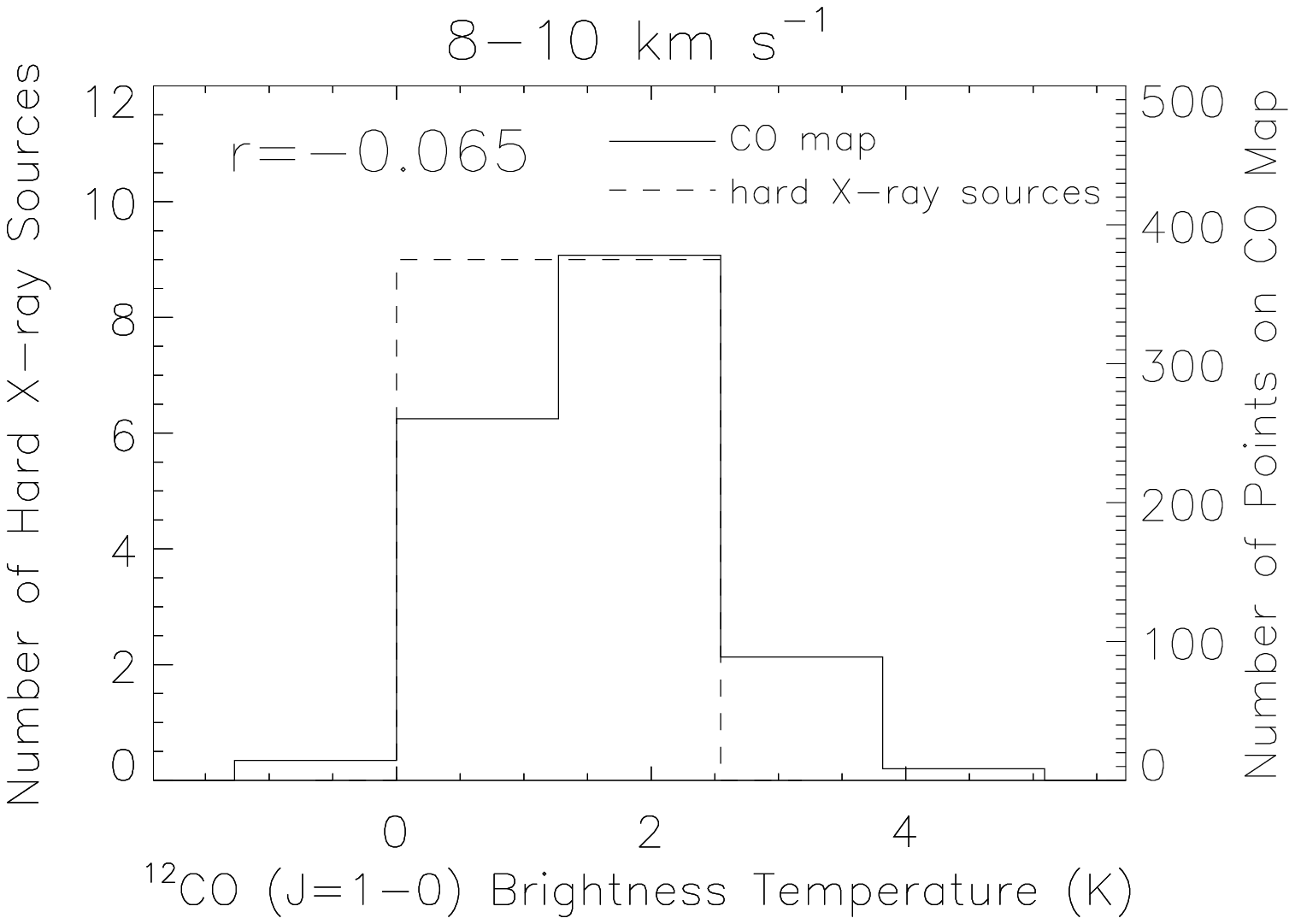}}

\centerline{\includegraphics[width=6.75cm]{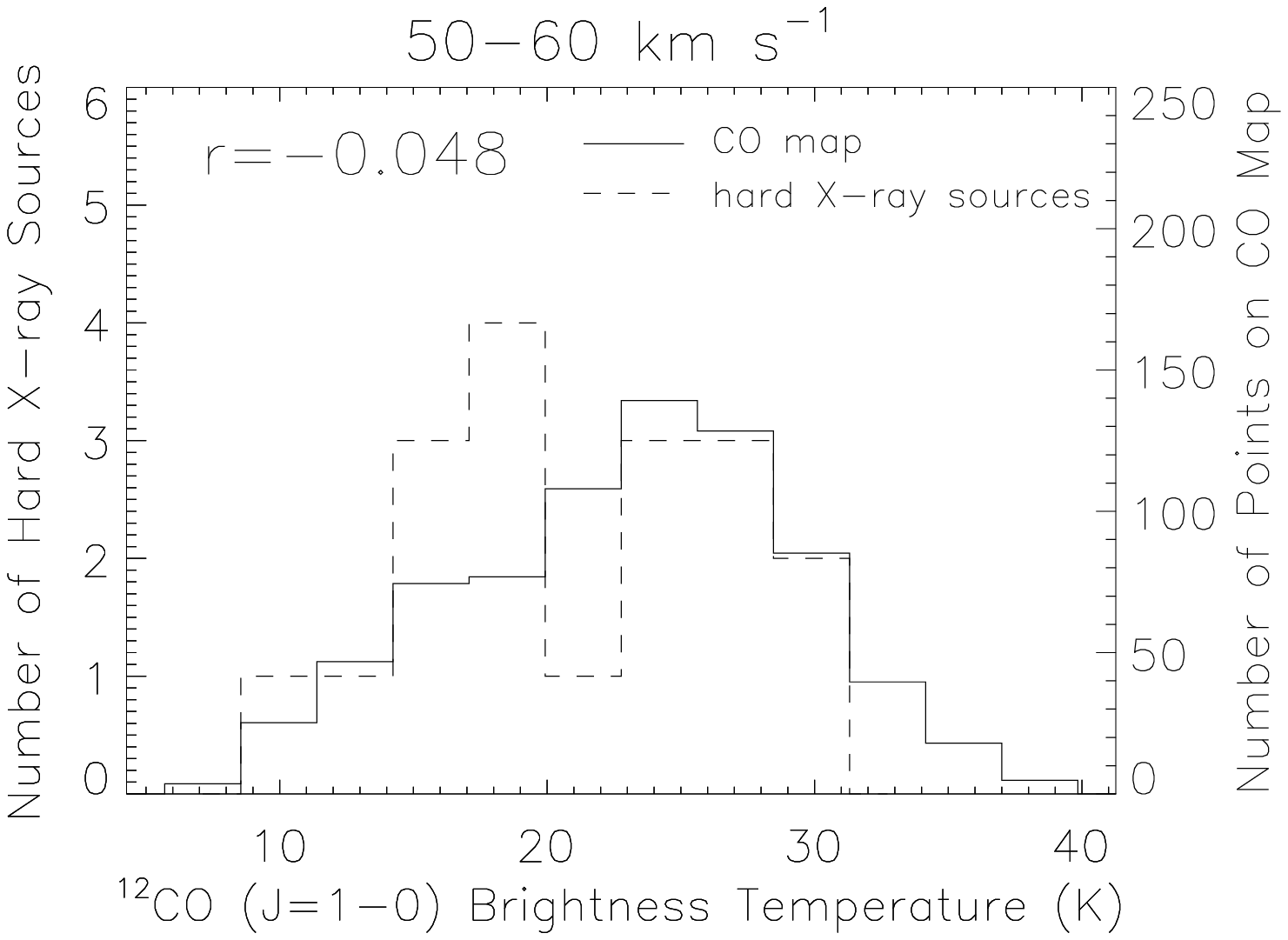}}

\centerline{\includegraphics[width=6.75cm]{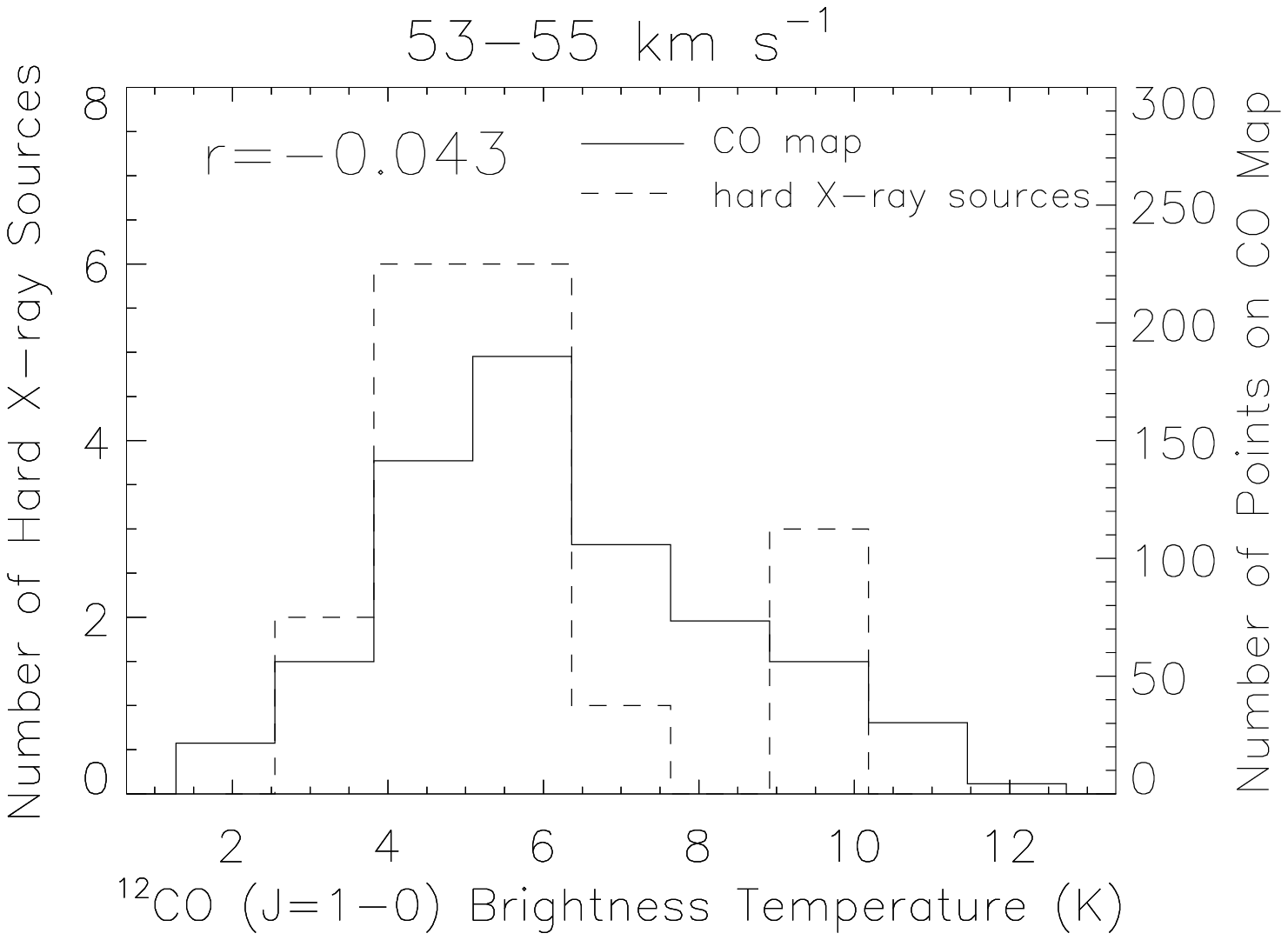}}

\centerline{\includegraphics[width=6.75cm]{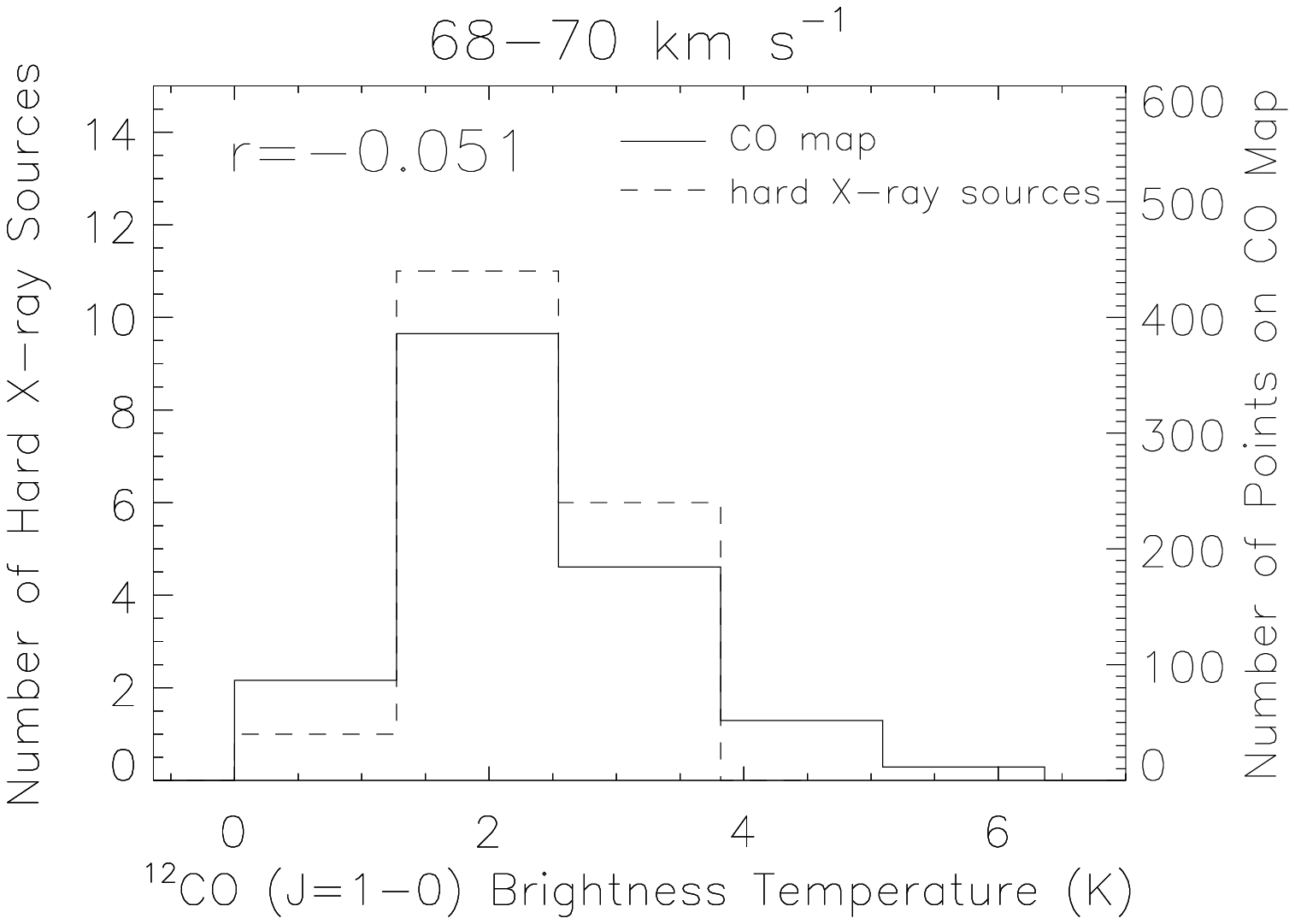}}

\caption{{\em From top to bottom:} as in Fig. \ref{histo80} but for 8-10, 50-60, 53-55 and 68-70 km s$^{-1}$ $^{12}$CO brightness temperature maps
}
\label{ohisto}
\end{figure}

To further strengthen this argument, we have also computed the histogram using the 8-10, 50-60, 53-55 and 68-70 km s$^{-1}$ $^{12}$CO brightness temperature maps, using the data of \citealt{zcs09}, and we show them in Fig. \ref{ohisto}. These velocity ranges are not associated to the remnant and therefore they should be in foreground. Indeed, the plots show that the source histogram either has the peak at the same brightness temperature of the cloud histogram, or at lower brightness level, indicating little or negative correlation.

We have also calculated the Pearson's correlation coefficient $r$ between the CO image and the hard X-ray sources image (an artificial image with the pixel values of one at the source position and zero elsewhere). This coefficient is 1 if the images are identical, $>0$ if the two images are correlate, 0 if they are completely uncorrelated, and $<0$ if they are anti-correlated. The correlation coefficient $r$ is shown in the upper-left corner of each image. The $80-81$ km s$^{-1}$ histogram is the only one with positive $r$, i.e. the only one indicating a possible correlation between the sources and the $^{12}$CO map.

\end{document}